\documentclass[sigconf]{acmart}

\renewcommand\footnotetextcopyrightpermission[1]{} 
\usepackage{booktabs} 
\usepackage{multirow}
\usepackage{subcaption}
\usepackage{caption}
\usepackage{setspace}
\usepackage{amsmath}
\usepackage[english]{babel}
\definecolor{mygray}{gray}{0.5}
\usepackage{float}
\usepackage{flushend}
\usepackage{mathtools}

\definecolor{midnightgreen}{rgb}{0.0, 0.29, 0.33}
\definecolor{orange}{RGB}{255,127,0}

\definecolor{Fern Green}{RGB}{57, 87, 33}

\setcopyright{none} 

\author{Yifan Qiao}
\affiliation{Tsinghua University}
\email{qiaoyf15@mails.tsinghua.edu.cn}

\author{Chenyan Xiong}
\affiliation{Microsoft Research}
\email{Chenyan.Xiong@microsoft.com}

\author{Zhenghao Liu}
\affiliation{Tsinghua University}
\email{liu-zh16@mails.tsinghua.edu.cn}

\author{Zhiyuan Liu}
\affiliation{Tsinghua University}
\email{liuzy@tsinghua.edu.cn}

\begin{document}
\title{Understanding the Behaviors of BERT in Ranking}


\begin{abstract}
This paper studies the performances and behaviors of BERT in ranking tasks.
We explore several different ways to leverage the pre-trained BERT and fine-tune it on two ranking tasks: MS MARCO passage reranking and TREC Web Track ad hoc document ranking.
Experimental results on MS MARCO demonstrate the strong effectiveness of BERT in question-answering focused passage ranking tasks, as well as the fact that BERT is a strong interaction-based seq2seq matching model.
Experimental results on TREC show the gaps between the BERT pre-trained on surrounding contexts and the needs of ad hoc document ranking.
Analyses illustrate how BERT allocates its attentions between query-document tokens in its Transformer layers, how it prefers semantic matches between paraphrase tokens, and how that differs with the soft match patterns learned by a click-trained neural ranker.
\end{abstract}

\maketitle

\section{Introduction}
    
    

In the past several years, neural information retrieval (Neu-IR) research has developed several effective ways to improve ranking accuracy. \emph{Interaction-based} neural rankers soft match query-documents using their term interactions~\cite{guo2016deep}; \emph{Representation-based embeddings} capture relevance signals using distributed representations~\cite{K-NRM, zamani2017relevance}; large capacity networks learn relevance patterns using large scale ranking labels~\cite{K-NRM, dai2018convolutional, pang2017deeprank}.
These techniques lead to promising performances on various ranking benchmarks~\cite{pang2017deeprank, K-NRM, guo2016deep, dai2018convolutional}.

Recently, BERT, the pre-trained deep bidirectional Transformer, has shown strong performances on many language processing tasks~\cite{BERT}.
BERT is a very deep language model that is pre-trained on the surrounding context signals in
large corpora.
Fine-tuning its pre-trained deep network works well on many downstream sequence to sequence (seq2seq) learning tasks.
Different from seq2seq learning, previous Neu-IR research considers such surrounding-context-trained neural models not as effective in search as relevance modeling ~\cite{K-NRM, zamani2017relevance}.
However, on the MS MARCO passage ranking task, fine-tuning BERT and treating ranking as a classification problem outperforms existing Neu-IR models by large margins~\cite{nogueira2019passage}.

This paper studies the performances and properties of BERT in ad hoc ranking tasks.
We explore several ways to use BERT in ranking, as representation-based and interaction-based neural rankers, as in combination with standard neural ranking layers.
We study the behavior of these BERT-based rankers on two benchmarks:
the MS MARCO passage ranking task, which ranks answer passages for questions,
and TREC Web Track ad hoc task, which ranks ClueWeb documents for keyword queries.

Our experiments observed rather different performances of BERT-based rankers on the two benchmarks.
On MS MARCO, fine-tuning BERT significantly outperforms previous state-of-the-art Neu-IR models, and the effectiveness mostly comes from its strong cross question-passage interactions.
However, on TREC ad hoc ranking, BERT-based rankers, even further pre-trained on MS MARCO ranking labels, perform worse than feature-based learning to rank and a Neu-IR model pre-trained on user clicks in Bing log.


We further study the behavior of BERT through its learned attentions and term matches.
We illustrate that BERT uses its deep Transformer architecture to propagate information more globally on the text sequences through its attention mechanism, compared to interaction-based neural rankers which operate more individually on term pairs.
Further studies reveal that BERT focuses more on document terms that directly match the query. It is similar to the semantic matching behaviors of previous surrounding context-based seq2seq models, but different from the relevance matches neural rankers learned from user clicks.


\section{BERT Based Rankers}

This section describes the notable properties of BERT and how it is used in ranking.

\subsection{Notable Properties of BERT}
We refer readers to the BERT and Transformer papers for their details~\cite{BERT, transformer}. Here we mainly discuss its notable properties that influence its usage in ranking. 

\textbf{Large Capacity.} BERT uses standard Transformer architecture---multi-head attentions between all term pairs in the text sequence---but makes it very deep. Its main version, \texttt{BERT-Large}, includes
24 Transformer layers, each with 1024 hidden dimensions and 16 attention heads.
It in total has 340 million learned parameters, much bigger than typical Neu-IR models.

\textbf{Pretraining.} BERT learns from the surrounding context signals in Google News and Wikipedia corpora. 
It is learned using two tasks: the first predicts random missing words (15\%) using the rest of the sentence (Mask-LM); the second predicts whether two sentences appear next to each other. In the second task, the two sentences are concatenated to one sequence; a special token ``[SEP]'' marks the sequence boundaries. 
Its deep network is very resource consuming in training: \texttt{BERT-Large} takes four days to train on 64 TPUs and easily takes months on typical GPUs clusters.

\textbf{Fine Tuning.} End-to-end training BERT is unfeasible in most academic groups due to resource constraints. It is suggested to use the pre-trained BERT as a fine-tuning method~\cite{BERT}. BERT provides a ``[CLS]'' token at the start of the sequence, 
whose embeddings are treated as the representation of the text sequence(s), and suggests to add task-specific layers on the ``[CLS]'' embedding in fine-tuning.

\subsection{Ranking with BERT}
We experiment with four BERT based ranking models: \texttt{BERT (Rep)}, \texttt{BERT (Last-Int)}, \texttt{BERT (Mult-Int)}, and \texttt{BERT (Term-Trans)}.
All four methods use the pre-trained BERT to obtain the representation of the query $q$, the document $d$, or the concatenation of the two $qd$. In the concatenation sequence $qd$, the query and document are concatenated to one sequence with boundary marked by a marker token (``[SEP]'').

The rest of this section uses subscript $i$, $j$, or $cls$ to denote the tokens in $q$, $d$, or $qd$, and superscript $k$ to denote the layer of BERT's Transformer network: $k=1$ is the first layer upon word embedding and $k=24$ or ``last'' is the last layer.
For example, $\vec{qd}_{cls}^k$ is the embedding of the ``[CLS]'' token, in the $k$-th layer of BERT on the concatenation sequence $qd$.

\textbf{BERT (Rep)} uses BERT to represent $q$ and $d$:
\begin{align}
   \texttt{BERT (Rep)}(q, d) &= \cos(\vec{q}_{cls}^{\text{last}}, \vec{d}_{cls}^{\text{last}}). 
\end{align}
It first uses the last layers' ``[CLS]'' embeddings as the query and document \emph{representations}, and then calculates the ranking score via their cosine similarity (cos).
Thus it is a \emph{representation-based} ranker.

\textbf{BERT (Last-Int)} applies BERT on the concatenated $qd$ sequence:
\begin{align}
\texttt{BERT (Last-Int)}(q, d) &= w^T \vec{qd}_{cls}^{\text{last}}.
\end{align}
It uses the last layer's ``[CLS]'' as the matching features and combines them linearly with weight $w$.
 It is the recommended way to use BERT~\cite{BERT} and is first applied to MARCO passage ranking by Nogueira and Cho~\cite{nogueira2019passage}.
The ranking score from \texttt{BERT (Last-Int)} includes all term pair interactions between the query and document via its Transformer's cross-match attentions~\cite{transformer}.
Thus it is an \emph{interaction-based} ranker.

\textbf{BERT (Mult-Int)} is defined as:
\begin{align}
\texttt{BERT (Mult-Int)}(q, d) &= \sum_{1 \leq k \leq 24} (w_{Mult}^k)^T \vec{qd}_{cls}^k.
\end{align}
It extends \texttt{BERT (Last-Int)} by adding the matching features $\vec{qd}_{cls}^k$ from all BERT's layers, to study whether different layers of BERT provide different information.

\textbf{BERT (Term-Trans)} adds a neural ranking network upon BERT, to study the performance of their combinations:
\begin{align}
s^k(q, d) &= \text{Mean}_{i,j}(\cos(\text{relu}(P^k \vec{q}_i^k), \text{relu}(P^k \vec{d}_j^k))) \\
\texttt{BERT }& \texttt{(Term-Trans)}(q, d) = \sum_k w_{trans}^k s^k(q,d).
\end{align}
It first constructs the translation matrix between query and document, using the cosine similarities between the projections of their contextual embeddings. Then it combines the translation matrices from all layers using mean-pooling and linear combination.

All four BERT based rankers are fine-tuned from the pre-trained BERT-Large model released by Google.
The fine-tuning uses classification loss, i.e., to classify whether a query-document pair is relevant or not, following the prior research~\cite{nogueira2019passage}. We experimented with pairwise ranking loss but did not observe any difference.

\section{Experimental Methodologies}
\noindent
\textbf{Datasets.} Our experiments are conducted on \texttt{MS MARCO} passage reranking task and TREC Web Track ad hoc tasks with \texttt{ClueWeb} documents.

\texttt{MS MARCO} includes question-alike queries sampled from Bing search log and the task is to rank candidate passages based on whether the passage contains the answer for the question\footnote{http://msmarco.org}. 
It includes 1,010,916 training queries and a million expert annotated answer passage relevance labels.
We follow the official train/develop split, and use the given ``Train Triples Small'' to fine-tune BERT. 

\texttt{ClueWeb} includes documents from ClueWeb09-B and queries from TREC Web Track ad hoc retrieval task 2009-2012. In total, 200 queries with relevance judgements are provided by TREC.
Our experiments follow the same set up in prior research and use the processed data shared by their authors~\cite{dai2018convolutional}: the same 10-fold cross validation, same data pre-processing, and same top 100 candidate documents from Galago SDM to re-rank.

We found that the TREC labels alone are not sufficient to fine-tune BERT nor train other neural rankers to outperform SDM.
Thus we decided to first pre-train all neural methods on MS MARCO and then fine-tune them on ClueWeb.

\textbf{Evaluation Metrics.} MS MARCO uses MRR@10 as the official evaluation. 
Results on the \textbf{Dev}elop set re-rank top 100 from BM25 in our implementation.
Results on \textbf{Eval}uations set are obtained from the organizers and re-rank top 1000 from their BM25 implementation.
ClueWeb results are evaluated by NDCG@20 and ERR@20, the official evaluation metrics of TREC Web Track.

Statistical significance is tested by permutation tests with $p<0.05$, except on MS MARCO Eval where per query scores are not returned by the leader board.

\textbf{Compared Methods.} 
The BERT based rankers are compared with the following baselines:
\begin{itemize}
\item \texttt{Base} is the base retrieval model that provides candidate documents to re-rank. It is BM25 on MS MARCO and Galago-SDM on ClueWeb.
\item \texttt{LeToR} is the feature-based learning to rank. It is RankSVM on MS MARCO and Coordinate Ascent on ClueWeb. 
\item \texttt{K-NRM} is the kernel-based interaction-based neural ranker~\cite{K-NRM}.
\item \texttt{Conv-KNRM} is the n-gram version of \texttt{K-NRM}.
\end{itemize}
\texttt{K-NRM} and \texttt{Conv-KNRM} results on ClueWeb are obtained by our implementations and pre-trained on MS MARCO.
We also include \texttt{Conv-KNRM (Bing)} which is the same Conv-KNRM model but pre-trained on Bing clicks by prior research~\cite{dai2018convolutional}.
The rest baselines reuse the existing results from prior research. 
Keeping experimental setups consistent makes all results directly comparable.

\textbf{Implementation Details.} 
All BERT rankers are trained using Adam optimizer and learning rate 3e-6, except \texttt{Term-Trans} which trains the projection layer with learning rate 0.002.
On one typical GPU, the batch size is 1 at most; fine-tuning takes on average one day to converge. Convergence is determined by the loss on a small sample of validation data (MS MARCO) or the validation fold (ClueWeb).
In comparison, K-NRM and Conv-KNRM take about 12 hours to converge on MS MARCO and one hour on ClueWeb. On MS MARCO all rankers take about 5\% training triples to converge.

\begin{table*}[h]
\centering
\caption{Ranking performances. Relative performances in percentages are compared to \texttt{LeToR}, the feature-based learning to rank. Statistically significant improvements are marked by $\dagger$ (over \texttt{Base}), $\ddagger$ (over \texttt{LeToR}), $\mathsection$ (over \texttt{K-NRM}), and $\mathparagraph$ (over \texttt{Conv-KNRM}).
Neural methods on ClueWeb are pre-trained on MS MARCO, except \texttt{Conv-KNRM (Bing)} which is trained on user clicks.
  \label{tab:msmarco}}
\begin{tabular}{l|lc|lc|lc|lc}
 \hline \hline
 & \multicolumn{4}{c|}{\bf{MS MARCO Passage Ranking}} & \multicolumn{4}{|c}{\bf{ClueWeb09-B Ad hoc Ranking}} \\ \hline
\bf{Method} & \multicolumn{2}{c|}{\bf{MRR@10 (Dev)}}& \multicolumn{2}{c|}{\bf{MRR@10 (Eval)}}
& \multicolumn{2}{|c}{\bf{NDCG@20}}& \multicolumn{2}{|c}{\bf{ERR@20}}
\\ \hline
\texttt{Base}  & ${0.1762}$ & $ -9.45\%  $
& $0.1649$ & $ +13.44\% $
 & ${0.2496}^{\mathsection }$ & $ -6.89\%  $
 & ${0.1387}$ & $ -14.25\%  $
\\
\texttt{LeToR}   & $0.1946$ & --
& ${0.1905}$ & --
 & $0.2681$ & --  & $0.1617$ & --
\\ \hline
\texttt{K-NRM}  & ${0.2100}^{\dagger \ddagger }$ & $ +7.92\%  $
& ${0.1982}$ & $ +4.04\% $
 & ${0.1590}$ & $ -40.68\%  $
 & ${0.1160}$ & $ -28.26\%  $
\\

\texttt{Conv-KNRM}  
 & ${0.2474}^{\dagger \ddagger \mathsection }$ & $ +27.15\%  $
 & ${0.2472}$ & $ +29.76\% $
 & ${0.2118}^{\mathsection }$ & $ -20.98\%  $
 & ${0.1443}^{\mathsection }$ & $ -10.78\%  $  \\

\texttt{Conv-KNRM (Bing)} & n.a. & n.a. 
& n.a. & n.a. 
& ${0.2872}^{\dagger \ddagger \mathsection \mathparagraph }$ & $ +7.12\% $
& ${0.1814}^{\dagger \ddagger \mathsection \mathparagraph }$ & $ +12.18\% $ \\

\texttt{BERT (Rep)}
 & ${0.0432}$ & $ -77.79\%  $
 & ${0.0153}$ & $ -91.97\%  $ 
 & ${0.1479}$ & $ -44.82\%  $
 & ${0.1066}$ & $ -34.05\%  $
\\


\texttt{BERT (Last-Int)}
 & ${0.3367}^{\dagger \ddagger \mathsection \mathparagraph }$ & $ +73.03\%  $
 & ${0.3590}$ & $ +88.45\%  $ 
 & ${0.2407}^{\mathsection \mathparagraph }$ & $ -10.22\%  $
 & ${0.1649}^{\dagger \mathsection \mathparagraph }$ & $ +2.00\%  $
\\

\texttt{BERT (Mult-Int)}
 & ${0.3060}^{\dagger \ddagger \mathsection \mathparagraph }$ & $ +57.26\%  $
 & ${0.3287}$ & $ +72.55\%  $ 
 & ${0.2407}^{\mathsection \mathparagraph }$ & $ -10.23\%  $
 & ${0.1676}^{\dagger \mathsection \mathparagraph }$ & $ +3.64\%  $
\\

\texttt{BERT (Term-Trans)}
 & ${0.3310}^{\dagger \mathsection \mathparagraph }$ & $ +70.10\%  $
 & ${0.3561}$ & $ +86.93\%  $ 
 & ${0.2339}^{\mathsection \mathparagraph }$ & $ -12.76\%  $
 & ${0.1663}^{\dagger \mathsection \mathparagraph }$ & $ +2.81\%  $
\\  \hline \hline
    \end{tabular}
  
\end{table*}

\begin{figure*}[h]
    \centering
    \includegraphics[width=0.9\columnwidth]{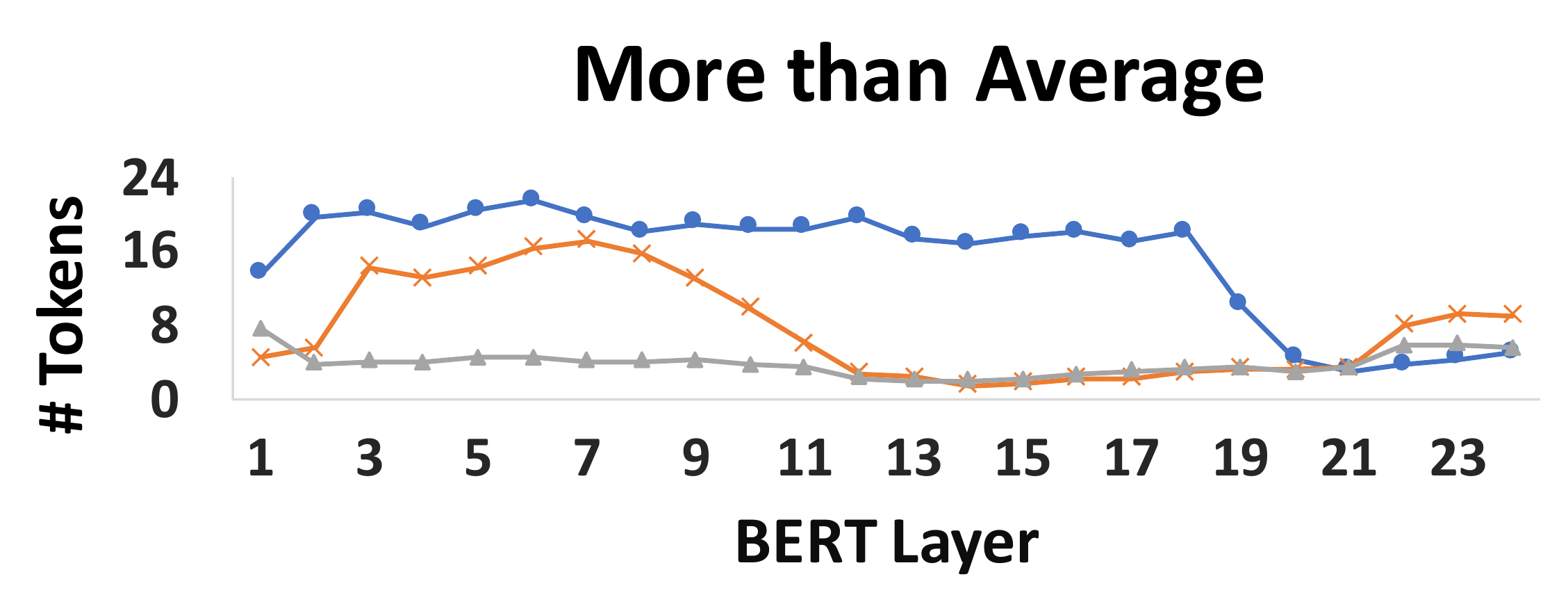}
    \includegraphics[width=0.9\columnwidth]{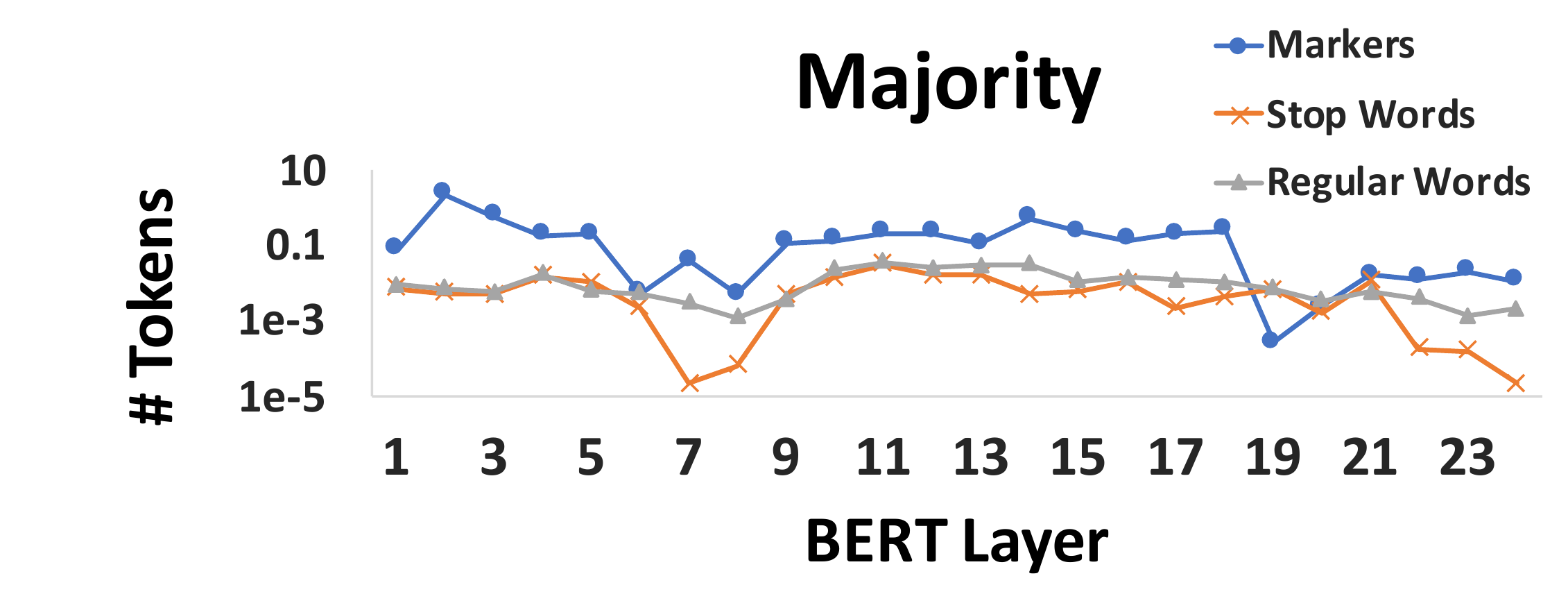}
    \caption{The attentions to Markers, Stopwords, and Regular Words in \texttt{BERT (Last-Int)}. X-axes mark layer levels from shallow (1) to deep (24). Y-axes are the number of tokens sending More than Average or Majority attentions to each group.
    \label{fig:att}
    }
    \label{fig:attention}
\end{figure*}

\section{Evaluations and Analyses}
This section evaluates the performances of BERT-based rankers and studies their behaviors.

\subsection{Overall Performances}
Table~\ref{tab:msmarco} lists the evaluation results on MS MARCO (left) and ClueWeb (right).
BERT-based rankers are very effective on MS MARCO: All interaction-based BERT rankers improved \texttt{Conv-KNRM}, a previous state-of-the-art, by 30\%-50\%.
The advantage of BERT in MS MARCO lies in the cross query-document attentions from the Transformers: \texttt{BERT (Rep)} applies BERT on the query and document individually and discard these cross sequence interactions, and its performance is close to random.
BERT is an \emph{interaction-based} matching model and is not suggested to be used as a representation model.

The more complex architectures in \texttt{Multi-Int} and \texttt{Term-Trans} perform worse than the simplest \texttt{BERT (Last-Int)}, even with a lot of MARCO labels to fine-tune.
It is hard to modify the pre-trained BERT dramatically in fine-tuning.
End-to-end training may make modifying pre-trained BERT more effective, but that would require more future research in how to make BERT trainable in accessible computing environments.

BERT-based rankers behave rather differently on ClueWeb.
Although pre-trained on large corpora and then on MARCO ranking labels, none of BERT models significantly outperforms \texttt{LeToR} on ClueWeb. 
In comparison, \texttt{Conv-KNRM (Bing)}, the same Conv-KNRM model but pre-trained on Bing user clicks~\cite{dai2018convolutional}, performs the best on ClueWeb, and much better than \texttt{Conv-KNRM} pretrained on MARCO labels.
These results demonstrate that MARCO passage ranking is closer to seq2seq task because of its question-answering focus, and BERT's surrounding context based pre-training excels in this setting. In comparison, TREC ad hoc tasks require different signals other than surrounding context: 
pre-training on user clicks is more effective than on surrounding context based signals.

\subsection{Learned Attentions}

This experiment illustrates the learned attention in BERT, which is the main component of its Transformer architecture.

Our studies focus on MS MARCO and \texttt{BERT (Last-Int)}, the best performing combination in our experiments, and randomly sampled 100 queries from MS MARCO Dev.
We group the terms in the candidate passages into three groups: Markers (``[CLS]'' and ``[SEP]''), Stopwords, and Regular Words. 
The attentions allocated to each group is shown in Figure~\ref{fig:att}.

The markers received most attention.
Removing these markers decreases the MRR by $15\%$: BERT uses them to distinguish the two text sequences.
Surprisingly, the stopwords received as much attention as non-stop words, but removing them has no effect in MRR performances. BERT learned these stopwords not useful and dumps redundant attention weights on them.

As the network goes deeper, less tokens received the majority of other tokens attention: the attention spreads more on the whole sequence and the embeddings are contextualized.
However, this does not necessarily lead to more global matching decisions, as studied in the next experiment.

\begin{figure}[h]
    \centering
    \begin{subfigure}{0.23\textwidth}
    \centering
    \includegraphics[width=\textwidth]{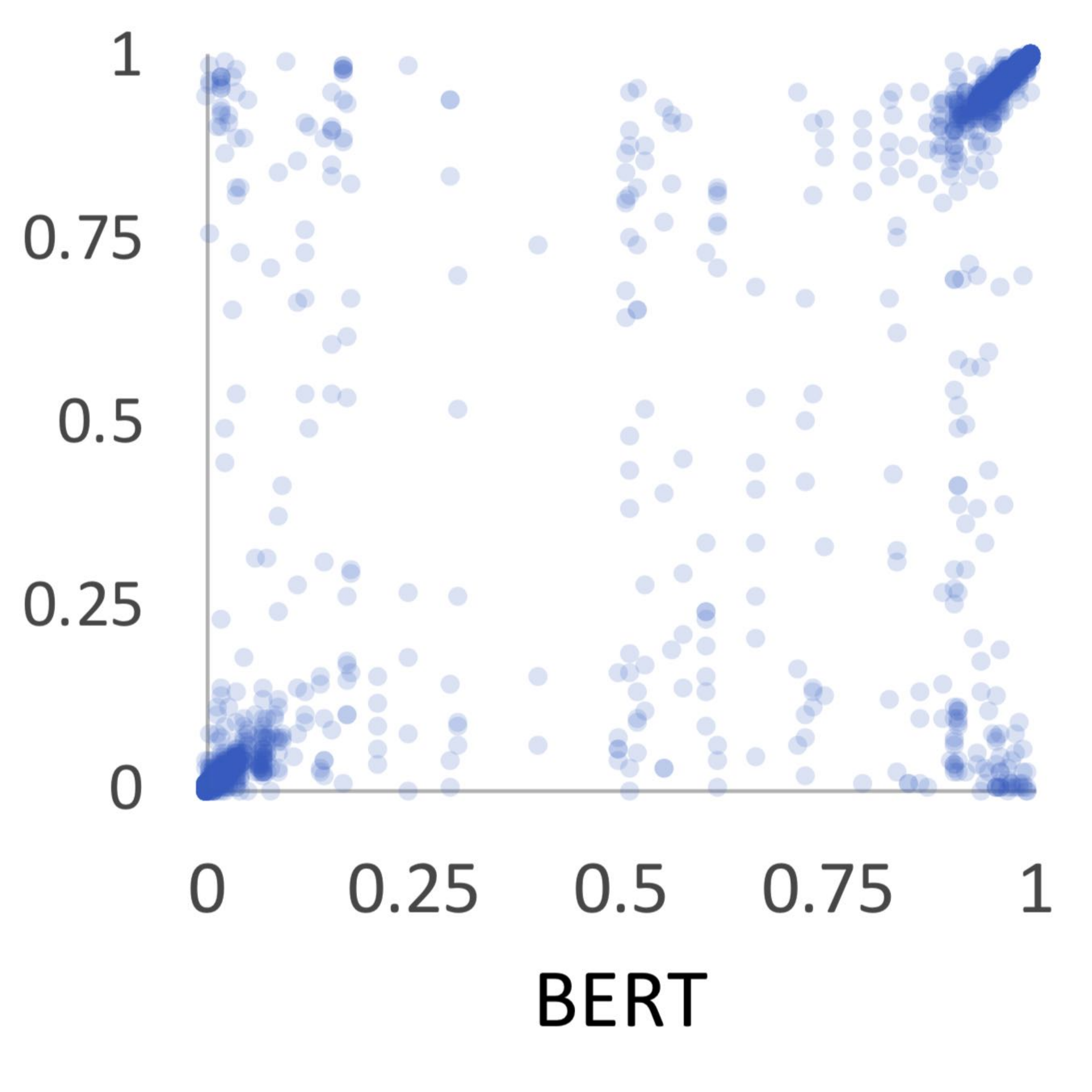}
    \end{subfigure}
    \begin{subfigure}{0.23\textwidth}
    \centering
    \includegraphics[width=\textwidth]{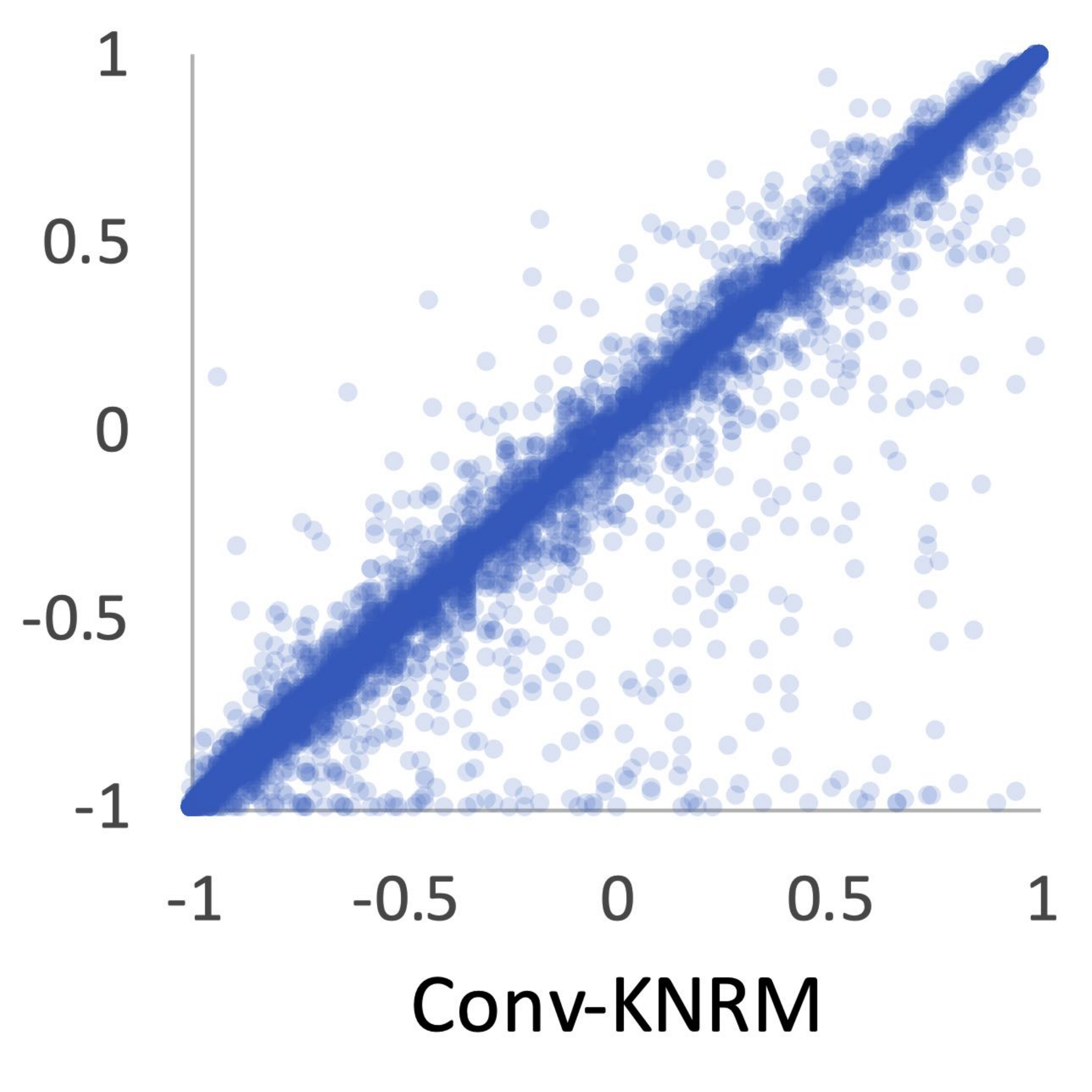}
    \end{subfigure}
    \caption{Influences of removing regular terms in {BERT (Last-Int)} and {Conv-KNRM} on MS MARCO. Each point corresponds to one query-passage pair with a random regular term removed from the passage. X-axes mark the original ranking scores and Y-axes are the scores after term removal.\label{fig:remove_select}}
\end{figure}

\subsection{Learned Term Matches}

        %
This experiment studies the learned matching patterns in \texttt{BERT (Last-Int)} and compares it to \texttt{Conv-KNRM}. The same MS MARCO Dev sample from last experiment is used.

We first study the influence of a term by comparing the ranking score of a document with and without the term. For each query-passage pair, we randomly remove a non-stop word, calculate the ranking score using \texttt{BERT (Last-Int)} or \texttt{Conv-KNRM}, and plot it w.r.t the original ranking score in Figure~\ref{fig:remove_select}.

Figure~\ref{fig:remove_select} illustrates two interesting behaviors of BERT.
First, it assigns more extreme ranking scores: most pairs receive either close to 1 or 0 ranking scores in BERT, while the ranking scores in Conv-KNRM are more uniformly distributed.
Second, there are a few terms in each document that determine the majority of \texttt{BERT}'s ranking scores; removing them significantly changes the ranking score---drop from 1 to near 0, while removing the majority of terms does not matter much in \texttt{BERT}---most points are grouped in the corners.
It indicates that BERT is well-trained from the large scale pre-training.
In comparison, terms contribute more evenly in \texttt{Conv-KNRM}; removing single term often varies the ranking scores of \texttt{Conv-KNRM} by some degree, shown by the wider band near the diagonal in Figure~\ref{fig:remove_select}, but not as dramatically as in \texttt{BERT}.

We manually examined those most influential terms in \texttt{BERT (Last-Int)} and \texttt{Conv-KNRM}. Some examples of those terms are listed in Table~\ref{tab:eg}.
The exact match terms play an important role in \texttt{BERT (Last-Int)}; we found many of the influential terms in BERT are those appear in the question or close paraphrases.
\texttt{Conv-KNRM}, on the other hand, prefers terms that are more loosely related to the query in search~\cite{dai2018convolutional}. For example, on MS MARCO, it focuses more on the terms that are the role of milk in macchiato (``visible mark''), the show and the role Sinbad played (``Cosby'' and ``Coach Walter''), and the task related to Personal Meeting ID (``schedule'').


These observations suggest that, BERT's pre-training on surrounding contexts favors text sequence pairs that are closer in their semantic meanings. 
It is consistent with previous observations in Neu-IR research, that such surrounding context trained models are not as effective in TREC-style ad hoc document ranking for keyword queries~\cite{K-NRM, dai2018convolutional, zamani2017relevance}.

\begin{table}[h]
    \centering
        \caption{Example of most influential terms in MS MARCO passages in BERT and Conv-KNRM.\label{tab:eg}}
    \begin{tabular}{l|c}
    \hline \hline
    \multicolumn{2}{l}{Query: ``What is a macchiato coffee drink''} \\ \hline
    \texttt{BERT (Last-Int)}     & macchiato,  coffee\\
    \texttt{Conv-KNRM}      &  visible mark\\ \hline
    \multicolumn{2}{l}{Query: ``What shows was Sinbad on''} \\ \hline
    \texttt{BERT (Last-Int)}     & Sinbad\\
    \texttt{Conv-KNRM}      &  Cosby, Coach Walter \\ \hline
    \multicolumn{2}{l}{Query: ``What is a PMI id''} \\ \hline
    \texttt{BERT (Last-Int)}     & PMI\\
    \texttt{Conv-KNRM}      &  schedule	a meeting \\ \hline \hline
    \end{tabular}
\end{table}

\section{Conclusions and Future Direction}
This paper studies the performances and behaviors of BERT in MS MARCO passage ranking and TREC Web Track ad hoc ranking tasks.
Experiments show that BERT is an interaction-based seq2seq model that effectively matches text sequences.
BERT based rankers perform well on MS MARCO passage ranking task which is focused on question-answering, but not as well on TREC ad hoc document ranking.
These results demonstrate that MS MARCO, with its QA focus, is closer to the seq2seq matching tasks where BERT's surrounding context based pre-training fits well, while on TREC ad hoc document ranking tasks, user clicks are better pre-training signals than BERT's surrounding contexts.

Our analyses show that BERT is a strong matching model with globally distributed attentions over the entire contexts.
It also assigns extreme matching scores to query-document pairs; most pairs get either one or zero ranking scores, showing it is well tuned by pre-training on large corpora.
At the same time, pre-trained on surrounding contexts, BERT prefers text pairs that are semantically close. This observation helps explain BERT's lack of effectiveness on TREC-style ad hoc ranking which is considered to prefer pretraining from user clicks than surrounding contexts.

Our results suggest the need of training deeper networks on user clicks signals.
In the future, it will be interesting to study how a much deeper model---as big as BERT---behaves compared to both shallower neural rankers when trained on relevance-based labels.

\bibliographystyle{ACM-Reference-Format}
\normalsize
\bibliography{citation}
\flushend 
\end{document}